
\input phyzzx
\REF\REVIEW{
For review, see, J. C. Le Guillou and J. Zinn-Justin eds.,
Large-Order Behavior of Perturbation Theory,
(North-Holland, Amsterdam 1990)}
\REF\BRE{
E. Br\'ezin, J. -C. Le. Guillou and J. Zinn-Justin,
{\sl Phys. Rev.\/} {\bf D15} (1977) 1558;\hfill\break
E. Br\'ezin, G. Parisi and J. Zinn-Justin,
{\sl Phys. Rev.\/} {\bf D16} (1977) 408.}
\REF\CHA{
S. Chadha and P. Olesen, {\sl Phys. Lett.\/} {\bf 72B} (1977)
87;\hfill\break
P. Olesen, {\sl Phys. Lett.\/} {\bf 73B} (1978) 327.}
\REF\KLE{
H. Kleinert, {\sl Phys. Lett.\/} {\bf B300} (1993) 261;\hfill\break
R. Karrlein and H. Kleinert, {\sl Phys. Lett.\/} {\bf A187} (1994) 133.}
\REF\ARV{
For a recent article, see, C. Arvanitis, H. F. Jones and C. S. Parker,
{\sl Phys. Rev.\/} {\bf D52} (1995) 3704.}
\REF\SEZ{
R. Seznec and J. Zinn-Justin, {\sl J.\ Math.\ Phys.\/} {\bf 20}
(1979) 1398.}
\REF\GUI{
J. C. Le Guillou and J. Zinn-Justin,
{\sl Ann.\ Phys.\ (N.Y.)\/} {\bf 147} (1983) 57.}
\REF\GKS{
R. Guida, K. Konishi, H. Suzuki, {\sl Ann. Phys. (N.Y.)\/} {\bf 241} (1995)
152;
\hfill\break
hep-th/9505084, to appear in {\sl Ann. Phys. (N.Y.)}.
}
\REF\HER{
I. W. Herbst and B. Simon, {\sl Phys. Rev. Lett.\/} {\bf 41} (1978) 67.}
\REF\BEN{
L. Benassi, V. Grecchi, E. Harrell and B. Simon,
{\sl Phys. Rev. Lett.\/} {\bf 42} (1979) 704;
{\sl Phys. Rev. Lett.\/} {\bf 42} (1979) 1430 (E).}
\REF\ALE{
M. H. Alexander, {\sl Phys. Rev.\/} {\bf 178} (1969) 34;\hfill\break
M. Hehenberger, H. V. McIntosh and E. Br\"andas, {\sl Phys. Rev.\/}
{\bf A10} (1974) 1494;\hfill\break
P. Froelich and E. Br\"andas, {\sl Phys. Rev.\/} {\bf A12} (1975) 1.}
\REF\MOR{See for example, P. M. Morse and H. Feshbach,
Methods of Theoretical Physics, p.~1676 (McGraw-Hill, New York, 1953)}
\REF\WU{
C. M. Bender and T. T. Wu, {\sl Phys. Rev.\/} {\bf 184} (1969) 1231.}
%
%
\overfullrule=0pt
\pubnum={IU-MSTP/5; hep-th/9512197}
\date={December 1995}
\titlepage
\title{The Hydrogen Atom in Strong Electric Fields:\break
Summation of the Weak Field Series Expansion}
\author{
Ken-ichi Hiraizumi, Yoshihisa Ohshima and Hiroshi Suzuki\foot{
e-mail: hsuzuki@mito.ipc.ibaraki.ac.jp}}
\address{
Department of Physics, Ibaraki University, Mito 310, Japan}
\abstract{
The order dependent mapping method, its convergence has recently been
proven for the energy eigenvalue of the anharmonic oscillator, is
applied to re-sum the standard perturbation series for Stark effect
of the hydrogen atom. We perform a numerical experiment up to the
fiftieth order of the perturbation expansion. A simple mapping suggested
by the analytic structure and the strong field behavior gives an
excellent agreement with the exact value for an intermediate range
of the electric field, $0.03\leq E\leq0.25$. The imaginary part of the
energy (the decay width) as well as the real part of the energy is
reproduced from the standard perturbation series.
}
\endpage

Quite often the perturbation series in quantum theory is not only
divergent but even non-Borel summable [\REVIEW]. Such a non-Borel
summability is sometimes known to have its physical origin. For quantum
mechanics with an unstable potential or degenerated potential minima
the tunneling (barrier penetration) effect gives the Borel singularity
[\BRE]. For a uniform electric field in QED, it can explicitly be shown
[\CHA] that the instability of the vacuum state causes the
non-Borel summability. One may even conjecture [\CHA] that the Borel
singularity in QED and QCD is an indication of the instability of the
perturbative vacuum.

The above observations in turn suggest a possibility that for certain case
the standard perturbation series has enough information on the quantum
tunneling (or the vacuum instability) that is usually regarded as
``non-perturbative'' effect.

Recently Kleinert [\KLE] has made an interesting observation that the
decay width (imaginary part of the energy eigenvalue) for an unstable
potential (anharmonic oscillator (AHO) with the negative coupling $g<0$)
can precisely be reproduced only by using the perturbation series
of the energy eigenvalue of AHO, hence explicitly implemented the
above possibility. His method is based on the so-called delta expansion
[\ARV] (variational perturbation method) that is a special case of the
order dependent mapping method (ODM) [\SEZ,\GUI]. The convergence of
the method for AHO has later been rigorously proven [\GKS]
for $|g|\geq g_0$ ($g_0$ is the radius of convergence of the strong
coupling expansion $g_0\sim0.1$). The proof itself [\GKS]
is applicable for other quantum systems if one has enough information on
the analyticity and the strong coupling behavior of the interested
quantity.

In this article, we take a simple quantum mechanical system for which
the perturbation expansion exhibits the Borel singularity on the
positive real axes and apply ODM to re-sum the standard perturbative expansion.
The physical origin of the non Borel summability in this system is
the instability of the ``vacuum.''

We consider the ground state energy of the hydrogen atom in a uniform
electric field (Stark effect),
$$
  \left(-{\hbar^2\over2m_e}\nabla^2-{e^2\over r}-eEz\right)\Psi
  ={\cal E}\Psi.
\eqn\one
$$
This system is unstable due to quantum tunneling (or spontaneous
ionization) and thus the energy eigenvalue of the quasi-ground state has
the imaginary part ($=-\Gamma/2$)\foot{In what follows,
the energy ${\cal
E}$ and the field strength $E$ are measured respectively by,
$2{\rm (Ry)}=m_ee^4/\hbar^2=27.21{\rm (eV)}$ and
$1{\rm (a.u.)}=m_e^2e^5/\hbar^4=5.142\times10^9{\rm (V/cm)}$.
Under a realistic condition in laboratory,
the maximum strength of the electric field is $E\sim2\times10^{-4}{\rm
(a.u.)}$, and practically is in the validity region of the standard
perturbation theory and WKB approximation. The aim of this article is
thus somewhat academic and to investigate how one can extract
information from a badly divergent (and non-Borel summable)
perturbation series.}.

On the other hand, the standard weak field expansion of the ``ground
state'' energy reads\foot{Since the perturbation hamiltonian has an
odd parity, only the even powers of the electric field survives in
the expansion for the ground state energy.}
$$
   {\cal E}(E)\sim\sum_{n=0}^\infty a_nE^{2n},
\eqn\two
$$
where the first several coefficients are
$$
\eqalign{
   &a_0=-1/2,\quad
    a_1=-9/4,\quad
    a_2=-3\,555/64
\cr
   &a_3=-2\,512\,779/512,\quad
    a_4=-13\,012\,777\,803/16\,384,
\cr
   &a_5=-25\,497\,693\,122\,265/131\,072.
\cr
}
\eqn\three
$$
(An economical way to compute the higher order coefficients
$a_n$ is summarized in Appendix.) The simple (truncated) sum of the
series is of course real for $E$ real. Note all the coefficients have
the same sign. In what follows we will apply ODM [\SEZ,\GUI] to this
standard weak field expansion and see how the exact energy eigenvalue
(that is complex) emerges.

As was emphasized in [\GKS], knowledge on the analytical property and
the strong coupling (field) behavior of the physical quantity are crucial
for applying ODM.
For the quasi-ground state of \one: i) The energy eigenvalue ${\cal E}(E)$
is analytic on the cut $E$ plane at least in the vicinity of the origin
and the cut is along positive real axis of $E$, as well as the negative axis,
due to the quantum tunneling [\HER]. ${\cal E}(E)$ is analytic on cut
$-E^2$ plane (at least in the vicinity of the origin), only the negative real
axis being the cut. ii) The discontinuity along the cut is given by the
WKB formula (for $E>0$)
$$
   {\rm Im}\,{\cal E}(E)=
   -{2\sqrt{2}\over1+\sqrt{2}}e^{\sqrt{2}-\pi/4}
   E^{-1}e^{-2/(3E)}(1+O(E))
   \sim-2E^{-1}e^{-2/(3E)}.
\eqn\addone
$$
The above two properties in turn give the large order behavior of the
perturbation coefficients in \two\ [\BEN]
$$
   a_n\sim-{4\over\pi}\left({3\over2}\right)^{2n+1}(2n)!.
\eqn\four
$$
Corresponding to the instability of the ground state, the series is
non-Borel summable. iii) The strong field behavior
[\BEN] is given by
$$
\eqalign{
   &{\rm arg}\,{\cal E}\sim-{\pi/3}+O((\ln E)^{-1}),
\cr
   &|{\cal E}|\sim 2^{-5/3}E^{2/3}(\ln E)^{2/3}
                   +{2^{4/3}\over3}E^{2/3}(\ln\ln E)(\ln E)^{-1/3}
                   +O(E^{2/3}(\ln E)^{-1/3}).
\cr
}
\eqn\fourteen
$$
This is an asymptotic expansion in contrast to the {\it convergent\/} strong
coupling expansion in AHO case.

In ODM [\SEZ,\GUI], one first map the coupling constant $E$ to
another variable $\lambda$ by some function $F(\lambda)$
$$
   E=\rho F(\lambda),\quad F(\lambda)=\lambda+O(\lambda^2),
\eqn\addtwo
$$
and factorize the original function with some prefactor function $f(\lambda)$
$$
   {\cal E}(E)=f(\lambda)\Psi(\lambda),
\eqn\addthree
$$
and expand $\Psi(\lambda)$ with respect to $\lambda$. The parameter of
the mapping $\rho$ is determined order by order by some condition.
The choice of $F(\lambda)$ and $f(\lambda)$ is crucial for the
convergence property of the method, and is constrained by the analytic
property and the strong field behavior of ${\cal E}(E)$ [\SEZ,\GUI,\GKS].

In the case of AHO [\SEZ,\GKS], the (known) best choice of the mapping
is equivalent to the delta expansion [\ARV] which starts from a
decomposition of the hamiltonian
$$
   H=-{1\over2}p^2+{\Omega^2\over2}q^2
   +\delta\left({\omega^2-\Omega^2\over2}q^2+{g\over4}q^4\right),
\eqn\addfour
$$
and the expansion with respect to $\delta$ (and finally $\delta=1$).
The variational parameter $\Omega$ (which corresponds to $\rho$ in ODM)
is determined order by order of the expansion by some condition.
It can be shown [\GKS] that the sequence obtained in this way converges to
the exact energy eigenvalue for a wide range of complex $g$.

As our first choice of the mapping in the present problem
we may therefore start from a decomposed hamiltonian:
$$
   H=-{1\over2}\nabla^2-{Z\over r}+\delta\left({Z-1\over r}-Ez\right),
\eqn\five
$$
where in the spirit of the delta expansion, we have introduced a variational
parameter $Z$. The expansion is done with respect to $\delta$ (and
finally $\delta=1$) and $Z$ is determined order by order by some
condition. As for AHO [\SEZ], it is easy to see the expansion is
equivalent to a mapping
$$
   E=\rho{\lambda\over(1-\lambda)^3},
\eqn\six
$$
and a factorization
$$
   {\cal E}(E)={1\over(1-\lambda)^2}\Psi(\lambda).
\eqn\seven
$$
The relation of parameters in both methods is given by (setting $\delta=1$)
$$
   \lambda=1-{1\over Z},\quad\rho={E\over Z^2(Z-1)}.
\eqn\eight
$$
The expansion of $\Psi(\lambda)$ up to $N$th order reads
$$
  \Psi_N(\lambda)=\sum_{n=0}^N b_n(\rho)\lambda^n
\eqn\nine
$$
where
$$
\eqalign{
   &b_0(\rho)=a_0,\quad b_1(\rho)=-2a_0,\quad b_2(\rho)=a_0+a_1\rho^2,
\cr
   &b_n(\rho)=
    \sum_{m=1}^{[n/2]}{(n+4m-3)!\over(n-2m)!(6m-3)!}a_m\rho^{2m},
   \quad{\rm for}\quad n\geq3.
\cr
}
\eqn\ten
$$
As the condition to fix $\rho$ order by order, we use [\SEZ,\GUI],
$$
   \Psi_N(\lambda)-\Psi_{N-1}(\lambda)=0,
\eqn\fifteen
$$
and pick up the real positive solution for $\rho$ [\SEZ,\GKS].
We may use the other condition such as [\ARV]
$\partial((1-\lambda)^{-2}\Psi_N)/\partial\rho=0$ as well. In the case
of AHO, the convergence property is known to be insensitive [\GKS]
to such a choice of the condition.

At first sight, the choice in \six\ and \seven\ seems to be well suited
to the strong field behavior of ${\cal E}$ \fourteen:
For $|E|$ large, $\lambda\sim1$ and $E\sim(1-\lambda)^{-3}$ from \six,
so the prefactor in \seven, $(1-\lambda)^{-2}\sim E^{2/3}$ reproduces
the correct power in \fourteen. The logarithm in \fourteen\ cannot be
reproduced in the present simple mapping.

We have done a numerical experiment based on the mapping \six\ and
\seven\ to $N=50$. For an actual calculation of the perturbation
coefficients $a_n$, see Appendix. For a wide range of $E$, we observed
a violating oscillating behavior such that
${\rm Re}\,{\cal E}(E)>0$ for $E=0.25$ from the first several orders.
The mapping \six\ and \seven\ completely fails.

The failure of the mapping \six\ and \seven\ that is equivalent to
a naive delta expansion like method \five, should clearly be related to the
incompatibility of the mapping \six\ and the analytic structure of ${\cal
E}(E)$. As was shown for AHO [\GKS] the mapping should be chosen to be
compatible with the analytic structure. In
this sense, the surprising success of the delta expansion for AHO was
somewhat accidental. In ODM on the other hand we have a wide freedom to
choose the form of the mapping and the decomposition\foot{Of course
it is quite possible that our choice of the variational parameter
in \five\ is not good enough to simulate the effect of the electric
field. One may introduce more complicated form of the variational
hamiltonian. Note however that for such a general choice, a
connection with the {\it standard\/} perturbation series is lost and hence
the advantage of the delta expansion is too.}. In fact according to
i) above the analogy with AHO case, for which the cut is only along
the negative real axis, is not hold.

This observation suggests a use of $-E^2$ instead of $E$ as
the fundamental variable for which the cut exists only along the
negative real axis (at least in the vicinity of the origin).
As the second choice therefor we take
$$
   -E^2=\rho{\lambda\over(1-\lambda)^\alpha},
\eqn\eleven
$$
and
$$
   {\cal E}={1\over(1-\lambda)^{\alpha/3}}\Psi(\lambda).
\eqn\twelve
$$
The form of the prefactor in \twelve\ has been chosen to reproduce the
$E^{2/3}$
behavior in \fourteen. Here we again abandon to reproduce the correct
logarithmic behavior in \fourteen\ and hence we expect the present mapping
fails for a large $E$ anyway [\GKS]. The coefficients $b_n(\rho)$ in
\nine\ now read
$$
   b_n(\rho)=\sum_{m=0}^n(-1)^m
   {\Gamma(n+(\alpha-1)m-\alpha/3)\over(n-m)!\Gamma(\alpha m-\alpha/3)}
   a_m\rho^m.
\eqn\thirteen
$$

For the value of $\alpha$, we do not have any criteria
at present (see below) and the choice is a sort of guesswork. We examined
$\alpha=3/2$, $2$, $5/2$, $3$, $7/2$, $4$ and $9/2$.
We have again done numerical experiments up to $N=50$. The $\rho$ is
determined again by the condition \fifteen\ and the real solution
for $\rho$ is taken.

For $\alpha=3/2$, we observed a tendency of a slow convergence to a
{\it wrong\/} answer, for example for $E=0.25$,
${\rm Re}\,{\cal E}\to-0.63$ and ${\rm Im}\,{\cal E}\to0.1$.
For $\alpha=2$, the sequence again converges but
to a wrong answer, namely ${\rm Re}\,{\cal E}\to -0.54$
and ${\rm Im}\,{\cal E}\to0.05$. $\alpha=4$ and $\alpha=9/2$ cases
do not work (slowly divergent). Among the above values, $\alpha=5/2$,
$\alpha=3$, $\alpha=7/2$ seems work and $\alpha=3$ seems to give the
fastest convergence. Therefore we will only report the result for
$\alpha=3$ in the following.

In Fig.~1, we have plotted the value of $\rho$ determined in this
way. A numerical fitting gives a scaling on the order $N$:
$$
   \rho\sim0.2851N^{-1.686}.
\eqn\addfive
$$

In Figs.~2a and 2b, the relative error to the exact value of the
energy (obtained by a numerical integration of the Schr\"odinger
equation [\ALE]) is depicted for a fixed value of the field strength
$E=0.25$ (2a for the real part and 2b for the imaginary part).
A tendency of the fast convergence to $N=50$ is observed.

Observing this convergent behavior, in Figs.~3a and 3b, we plotted
${\rm Re}\,{\cal E}(E)$ and ${\rm Im}\,{\cal E}(E)$ by ODM and the
exact values [\ALE] (full squares) in the intermediate range of the field
strength. We see a surprisingly good agreement for the imaginary part
as well as the real part.

As was mentioned above, for a strong field regime, we expect the method
fails because the present mapping can only reproduce the correct
power behavior in \fourteen\ but not the logarithmic corrections.
To see this we have plotted ${\rm Re}\,{\cal E}(E)$ and
${\rm Im}\,{\cal E}(E)$ by ODM and the asymptotic behavior \fourteen\
(broken line) in Figs.~4a and 4b.

On the other hand, the experience in AHO case [\GKS] suggest
the method also fails to reproduce the {\it imaginary\/} part
for the weak field region. In Table 1, the imaginary part obtained by
WKB approximation \addone\ (which is a good approximation within this
regime) and ODM are compared. It can be seen that ODM completely
fails for $E<0.025$\foot{For the {\it real\/} part of the energy,
ODM gives an excellent result in this weak field regime.}.

In summary we have numerically observed that, at least for an
intermediate range of the field strength, we can extract an accurate
value of the imaginary part from the standard perturbation series
by suitably choosing the mapping in ODM.

In spite of this success, the convergence proof of [\GKS] cannot
directly be applied to the present case:
First of all, $\alpha=3$ is a dangerous case in the view point of the
proof [\GKS]
because the negative real of $-E^2$ is mapped to $\lambda$ with
$|\lambda|<1$ which obstructs the direct application of the proof.
One have to extend the ``contour'' [\GKS] to the outside of $|\lambda|>1$
but this requires some knowledge on the analyticity on the higher
Riemann sheet of $-E^2$ that is lacking at present.

Secondly the prefactor $(1-\lambda)^{-1}$ (for $\alpha=3$)
in \twelve\ is not sufficient to make $\Psi(\lambda)$ finite
as $\lambda\to1$, because of the logarithmic correction in
\fourteen. The boundedness of $\Psi(\lambda)$ at $\lambda=1$
was another necessity condition for the convergence proof in [\GKS].

Finally the fact that the observed power in \addfive\
is larger than $-2$ is rather strange. Repeating the
analysis of [\GKS], the error of the method caused by the divergence of
the perturbation series ($R_N^{(A)}$ part in [\GKS]) is estimated as
$$
   |R_N^{(A)}|\sim
   {2\over\pi}\left({3\over2}\right)^{2N+1/2}
   {|\lambda|^N\over|1-\lambda|}\rho^N
   \Gamma(2N+1)(1+O(1/N))
$$
which diverges if the scaling \addfive\ retaines for $N\to\infty$ and
is in fact $\sim 156$ for $N=50$ and $E=0.25$ (The relative error
$\sim26300$\%). This clearly contradicts with the observed fast
convergence
in Figs~2a and 2b. The convergence mechanism in [\GKS] therefore
cannot be applied in the present case. We need some new explanation
such as a cancellation among $R_N^{(A)}$ part and $R_N^{(B)}$ part
in [\GKS]. The fact that $\Psi(\lambda)$ is not bounded at $\lambda=1$
also suggests this possibility.

Obviously more theoretical study is needed to explain the empirical
success of the mapping \eleven\ with $\alpha=3$. Nevertheless we believe
our observation gives another concrete support for the possibility
that one can in practice reproduce the exact property (tunneling or
decay in the present case) of the physical quantity only relying on the
standard perturbation series.

It is possible to reproduce the correct logarithmic behavior in \fourteen\
by using more complicated mapping such as the one in [\GUI]. Although
we do not try in this article, it is crucial if one wants an accurate
value for a large $E$.

We thank Prof.~T. Amano for useful information. The work of
H.S. is supported in part by Monbusho Grant-in-Aid
Scientific Research No.~07740199 and No.~07304029.

\appendix

It is well known [\MOR] that the Schr\"odinger equation \one\
has a separated form in the parabolic coordinate
$x=\sqrt{\lambda\mu}\cos\varphi$, $y=\sqrt{\lambda\mu}\sin\varphi$
and $z=(\lambda+\mu)/2$.
The problem is then reduced to one-dimensional and we may
apply the recursion formula technique in [\WU]. Setting
$$
   \Psi=(\xi\eta)^{m/2}e^{-(\xi+\eta)/2}F(\xi)G(\eta),
\eqn\appone
$$
($m$ is the magnetic quantum number) where
$$
   \xi=\sqrt{-2{\cal E}}\lambda,\quad\eta=\sqrt{-2{\cal E}}\mu,
\eqn\apptwo
$$
we have
$$
\eqalign{
   &\xi F''+(m+1-\xi)F'
   -\left({1\over2}m+{1\over2}-\sigma-{g\over4}\xi^2\right)F=0,
\cr
   &\xi G''+(m+1-\xi)G'
   -\left({1\over2}m+{1\over2}-\tau+{g\over4}\xi^2\right)G=0.
\cr
}
\eqn\appthree
$$
Here $\sigma$ and $\tau$ are the separation constants and
the ``coupling constant'' $g$ is defined by
$$
   \sigma+\tau={1\over\sqrt{-2{\cal E}}},\quad
   g={E\over(-2{\cal E})^{3/2}}.
\eqn\appfour
$$
In what follows we consider only the case $m=0$.

We define the ``perturbative expansion''
$$
   \sigma=\sum_{k=0}^\infty\sigma_kg^k,\quad
   F(\xi)=\sum_{j=0}^\infty F_j(\xi)g^j.
\eqn\appfive
$$
The zeroth order solution is given by $\sigma_0=1/2$ and $F_0(\xi)=1$.
The equation \appthree\ has a polynomial solution order by order in
$g$. Hence we further set
$$
   F_j(\xi)=\sum_{l=1}^{2j}f_{j,l}\xi^l,\quad{\rm for}\quad j\geq1.
\eqn\appsix
$$
Eq.~\appthree\ then becomes a set of recursion relations:
$$
   \sigma_j=-f_{j,1},
\eqn\appseven
$$
and
$$
   (l+1)^2f_{j,l+1}-lf_{j,l}+\sum_{k=1}^{j-1}\sigma_kf_{j-k,l}
   -{1\over4}f_{j-1,l-2}=0,
\eqn\appeight
$$
for $j\geq1$ ($f_{0,0}=1$ and $f_{0,l}=0$ for $l>0$).

We have numerically solved \appseven\ and \appeight\ by FORTRAN
in double precision up to $\sigma_{100}$. We then have used
a symbolic manipulation package MATHEMATICA to invert
\appfour\ (note $\tau=\sum_{k=0}^\infty(-1)^k\sigma_kg^k$)
for obtaining the perturbation coefficients $a_n$ in \two\
up to $a_{50}$.

To examine the quality of our numerical value for $a_n$,
we plotted the ratio of $a_n$ to the asymptotic behavior
\four\ in Fig.~5.

\refout
\vfill\eject
\centerline{\fourteenrm Table Caption}
\item{\rm Table\ 1}
Comperison of ODM with WKB formula \addone\ in the weak field region.
\bigskip
\centerline{\fourteenrm Figure Captions}
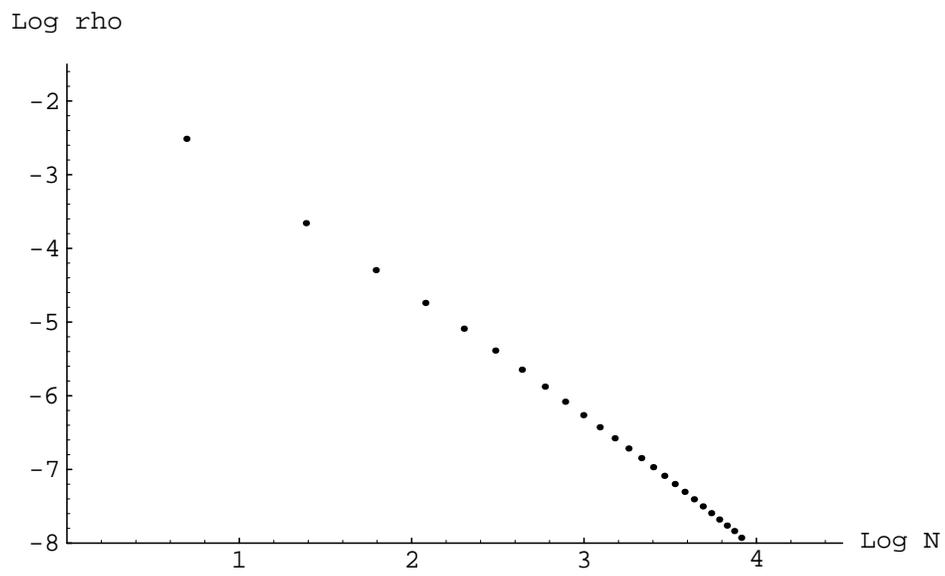
\item{\rm Fig.~1}
Scaling of $\rho$ with respect to the order of ODM $N$. $\rho$ is fixed
by the condition \fifteen.
\item{\rm Fig.~2}
The relative error of ODM \eleven\ with $\alpha=3$ (2a for the real
part, 2b for the imaginary part) with respect to the order $N$.
\item{\rm Fig.~3}
The real part (3a) and the imaginary part (3b) of the energy obtained by
ODM in the intermediate region of the field strength.
The full squares are the exact values [\ALE].
\item{\rm Fig.~4}
The real part (4a) and the imaginary part (4b) of the energy obtained by
ODM in the strong region of the field strength.
The asymptotic behavior \fourteen\ is depicted by broken lines.
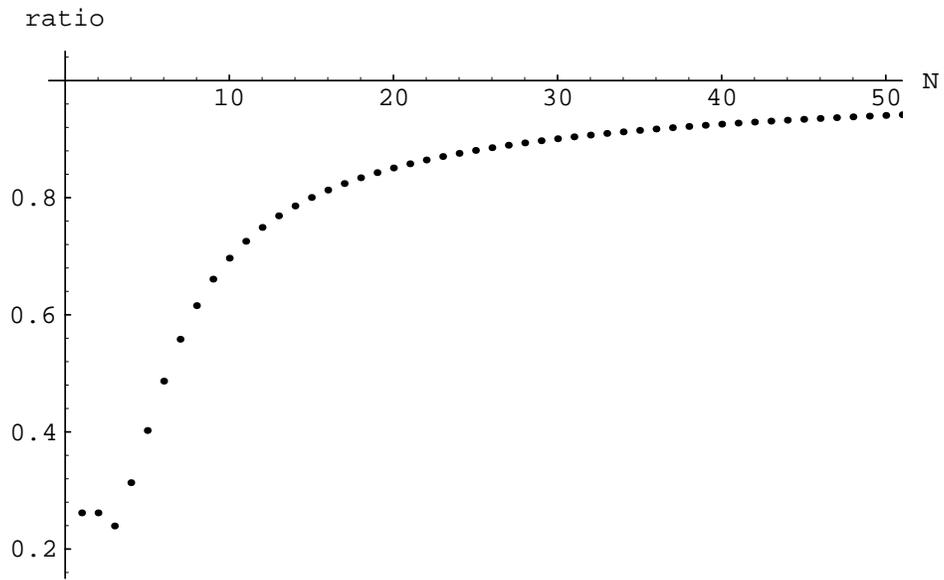
\item{\rm Fig.~5}
The ratio of the perturbative coefficients $a_n$ and the large order
behavior \four.
\vfill\eject
\null
\vfill
\centerline{
\vbox{
\halign{\quad#\hfil\quad&\quad#\hfil\quad&\quad#\hfil\quad\cr
$E$      & ${\rm Im}\,{\cal E}$ (WKB) & ${\rm Im}\,{\cal E}$ (ODM) \cr
$0.01$   & $-2.44847\times10^{-27}$ & $-1.38778\times10^{-16}$ \cr
$0.0125$ & $-1.20942\times10^{-21}$ & $+4.16334\times10^{-16}$ \cr
$0.015$  & $-7.30789\times10^{-18}$ & $+1.09635\times10^{-14}$ \cr
$0.0175$ & $-3.58320\times10^{-15}$ & $+7.32747\times10^{-14}$ \cr
$0.02$   & $-3.66731\times10^{-13}$ & $+3.27433\times10^{-13}$ \cr
$0.0225$ & $-1.32341\times10^{-11}$ & $-1.04966\times10^{-11}$ \cr
$0.025$  & $-2.30534\times10^{-10}$ & $-1.66882\times10^{-10}$ \cr
$0.0275$ & $-2.36688\times10^{-9}$  & $-1.66460\times10^{-9}$  \cr
$0.03$   & $-1.63588\times10^{-8}$  & $-1.11321\times10^{-8}$  \cr}
}
}
\vfill
\centerline{\fourteenrm Table. 1}
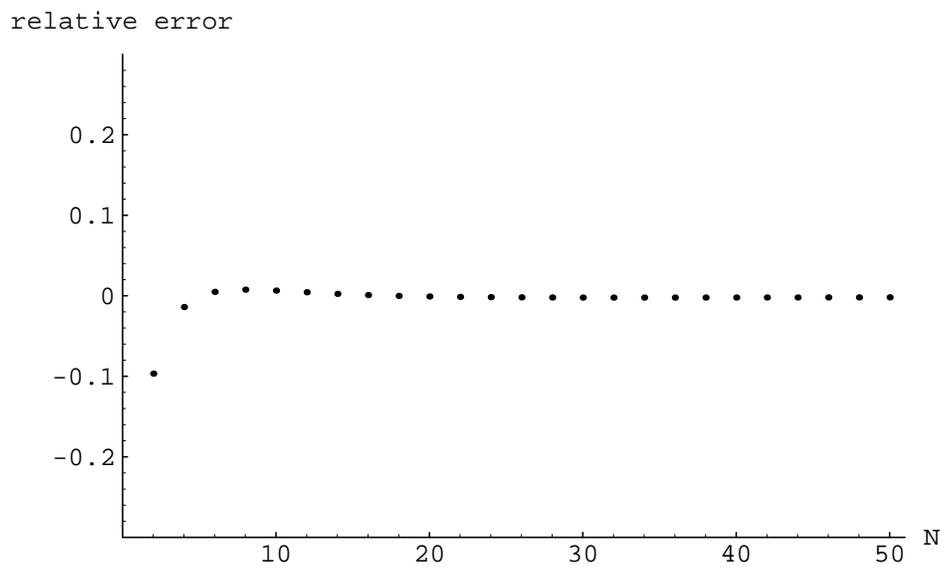
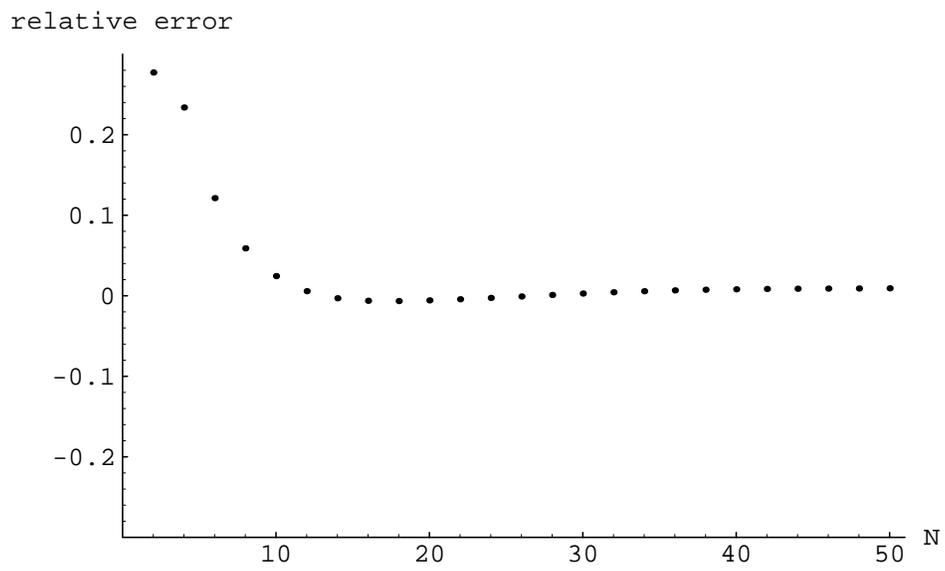
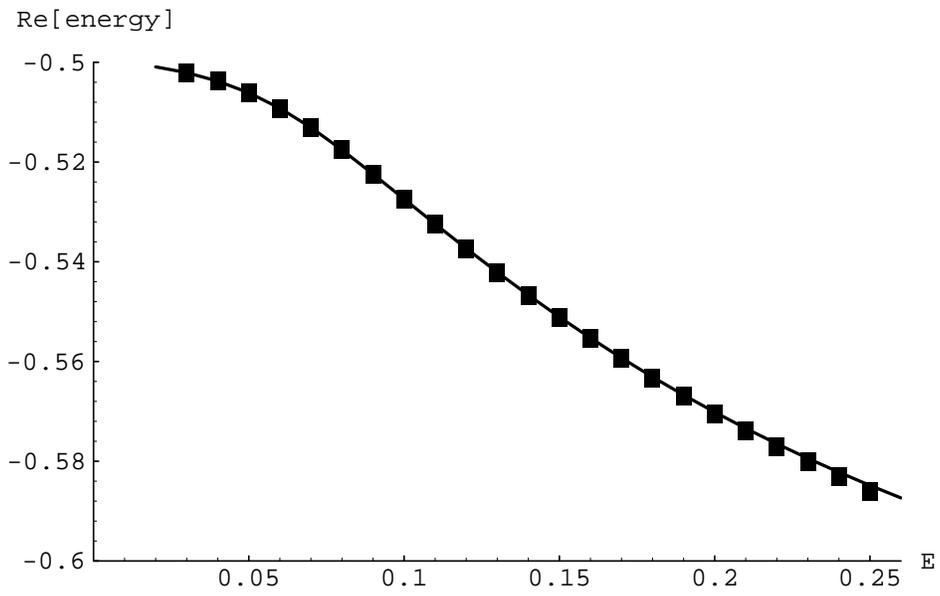
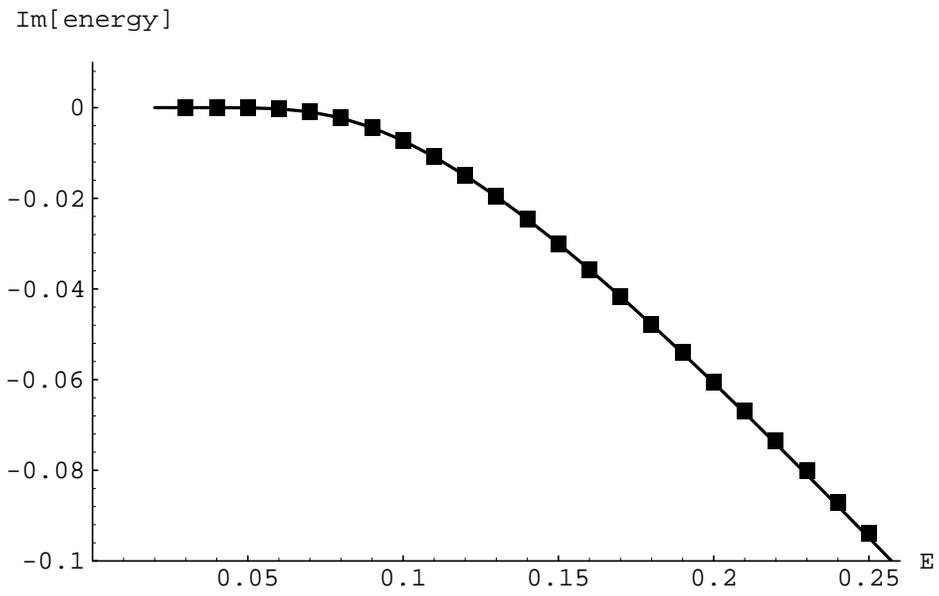
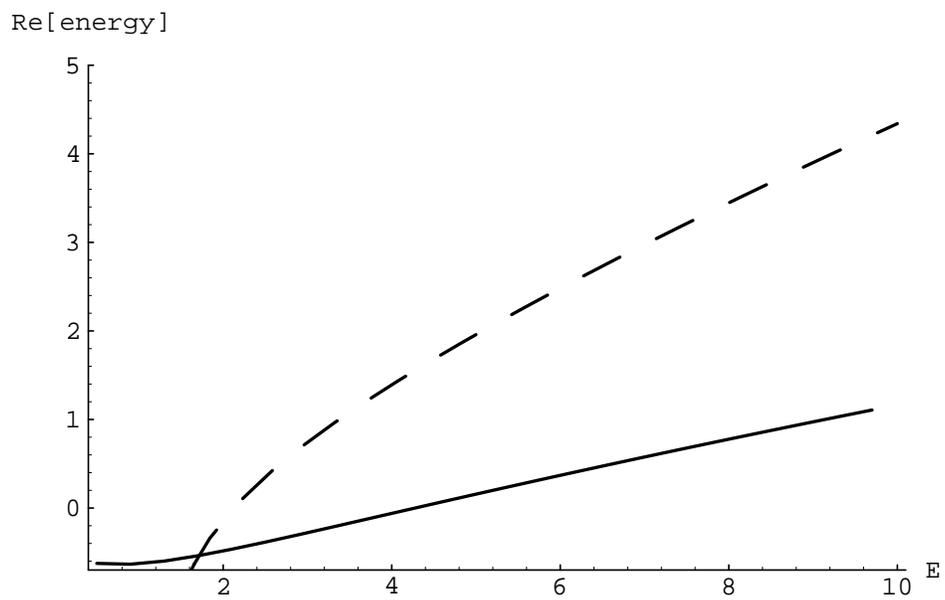
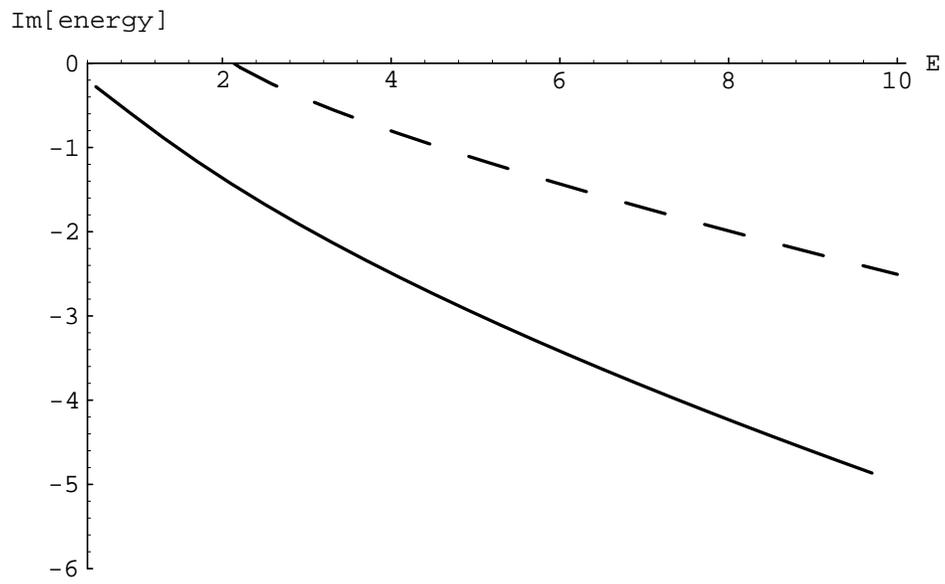
\bye